# Mobility Models for Vehicular Ad-hoc Network Simulation

Vaishali D. Khairnar
Symbiosis Institute of Technology
Pune

Dr. S.N.Pradhan
Institute of Technology
Nirma University, Ahmedabad

## ABSTRACT
One of the emerging applications that belong to ambient systems is to transparently and directly interconnect vehicles on roads, making an ad-hoc network that enables a variety of applications through distributed software's without the need of any fixed and dedicated infrastructure. The network as well as the embedded computers and sensors in the vehicle will be invisible to the driver, who will get the required services during his journey. New type of ad-hoc network is the Vehicular Ad-hoc Network (VANET), in which vehicles constitute the mobile nodes in the network. Due to the prohibitive cost of deploying and implementing such as system in a real world, most research work in VANET relies on simulations for evaluation purpose. The key concept for VANET simulations is a real world vehicular mobility model which will ensures conclusions drawn from simulation experiments will carry through to real world deployments. In this paper we present a tool SUMO, MOVE that allows users to easily generate real world mobility models for VANET simulations. MOVE tool is built on top of SUMO which is open source micro-traffic simulator. Output of MOVE is a real world mobility model and can used by NS-2 and qualnet simulator. In this paper we evaluate and compare ad-hoc routing performance for vehicular nodes using MOVE, which is using random waypoint model. The simulation results are obtained when nodes are moving according to a real world mobility model which is significantly different from that of the generally used random waypoint model.

## 1. INTRODUCTION
VANET communication is becoming an important and popular research topic in the area of wireless networking as well as in the automobile industries. Goal of this research is to develop a vehicular communication system which will enable quick and cost-efficient distribution of data for the benefit of passenger's safety and comfort.

It is crucial to test and evaluate protocol implementations in a real world environment, simulations are commonly used as a first step in the protocol development for VANET research. Several communications network simulator already exist to provide a platform for testing and evaluating network protocols, such as NS-2[1], OPNET [2] and Qualnet [3]. These tools are basically designed to provide generic simulation scenarios without being particularly tailored for specific applications in the transportation environment. Simulations play an important role in the area of transportation. Variety of simulation tools is available such as PARAMICS [4], CORSIM [5] and VISIM [6], MOVE [8], SUMO [9,20,12,22], NS-2, VanetMobiSim [7] etc which have been developed to analyze transportation scenarios at the micro- and macro-scale levels.

The most important parameters in simulating ad-hoc networks is the node mobility. It's important to use real world mobility model so that the results from the simulation correctly reflect the real-world performance of a VANET. E.g. a vehicle node is typically constrained to streets which are separated by buildings, trees or other objects. Such obstructions often increase the average distance between nodes as compared to an open-field environment [10, 11].

We will deploy a tool MOVE to which will provide facility for the users to generate real world mobility models for VANET simulations. MOVE tool is built on top of an open source micro-traffic simulator SUMO [9]. The output of MOVE is a mobility trace file that contains information of real-world vehicle movements which can be used by NS-2 or Qualnet. MOVE provides a set of GUI that allows the user to quickly generate real-world simulation scenarios without ant simulation scripts.

## 2. MOVE ARCHITECTURE
MOVE is implemented in Java and runs on top of an open-source micro-traffic simulator SUMO. Following syeps to implement MOVE.

- First implement java sdk 1.6 and NS-2 Version 2.34 on Fedora 13 operating system.
- Next implement XML parser 3.11, FOX toolkit, PROJ and GDAL on Fedora version 13 OS.
- Next configure SUMO with FOX, PROJ and GDAL.
- Execute MOVE by jar file.
- MOVE consists of two main components:-
- Map Editor

Vehicle Movement Editor, as shown in Fig.1,2 [13,14]
The Map Editor is used to create the road topology as shown in Example 1 Fig.3. Basically implementation provides three different ways to create the road map[17,18,19].
- ✓ The map can be manually created.
- ✓ Generated automatically
- ✓ Imported from existing real world maps such as Google maps.

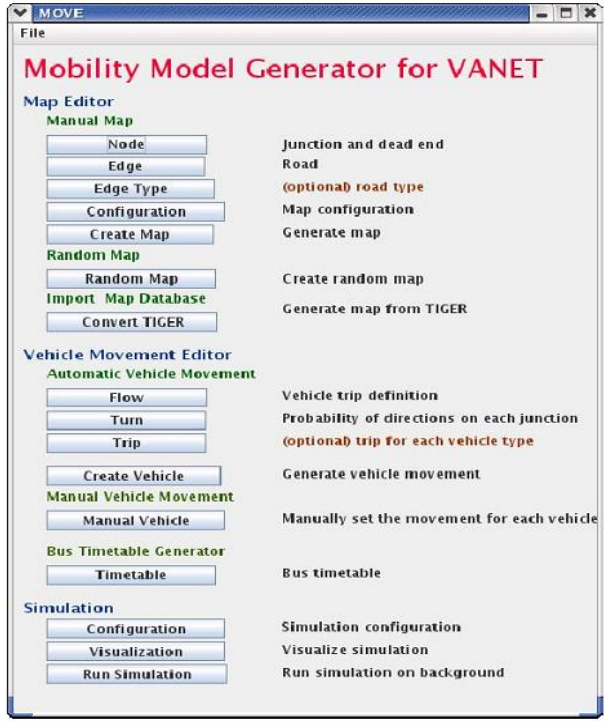

Fig.1

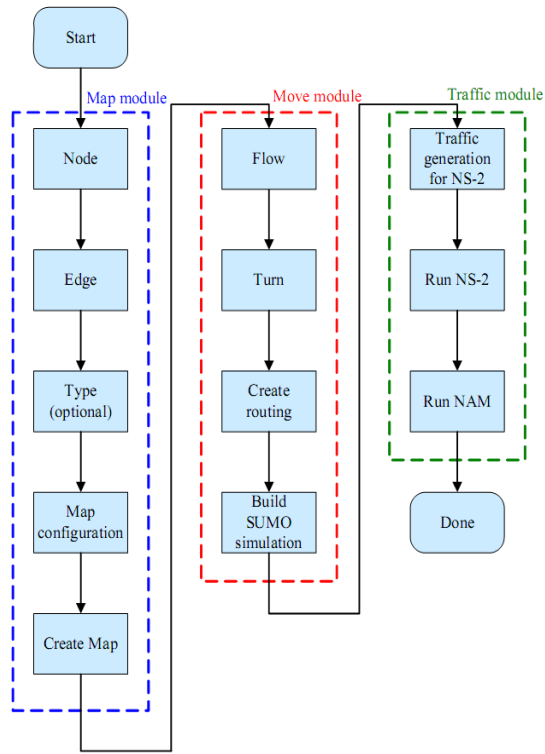

Fig.2.

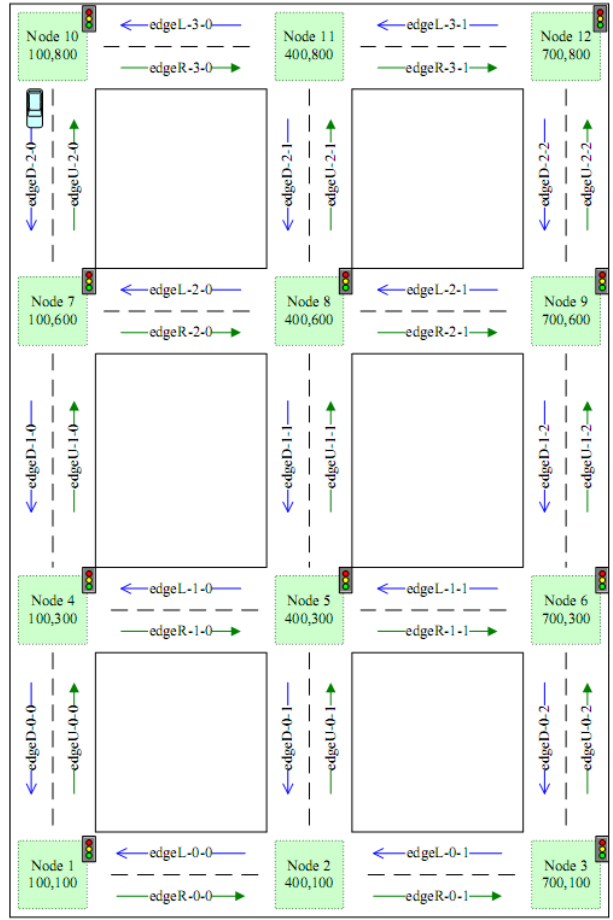

Fig.3. Example 1

Screen shots of Map Editor for Example 1 to create road topology as shown in Fig. 4.

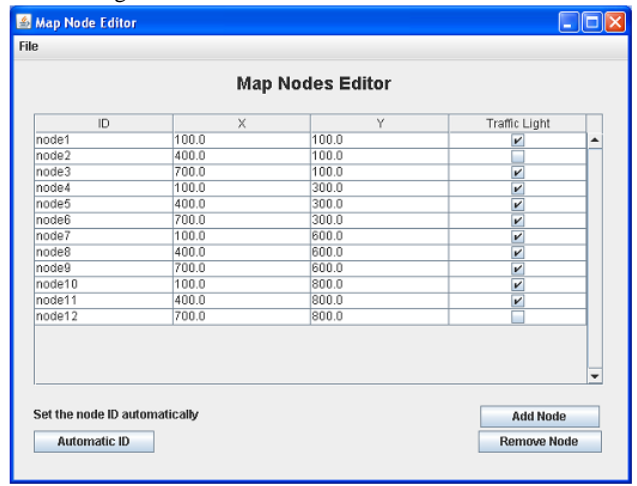

Fig. 4

The Vehicle Movement Editor allows user to specify the trips of vehicles and the route that each vehicle will take for one particular trip Fig.5. Three different methods to define the vehicle movements:-

9

- ✓ The vehicle movement patterns can be manually created
- ✓ Generated automatically Fig. 6.
- ✓ Specified based on a bus time table to simulate the movements of public transportations. Fig.1

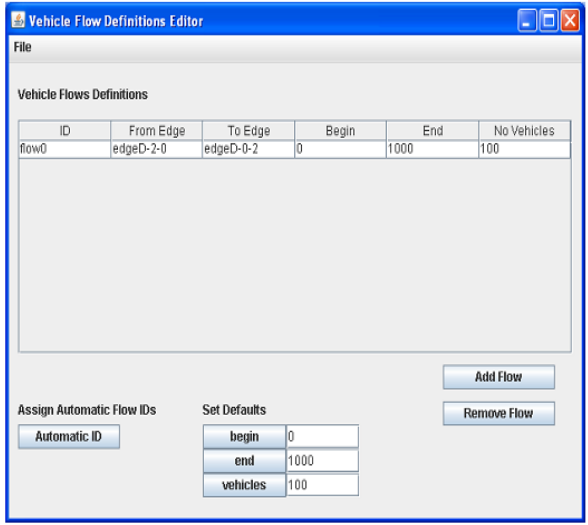

Fig. 5a

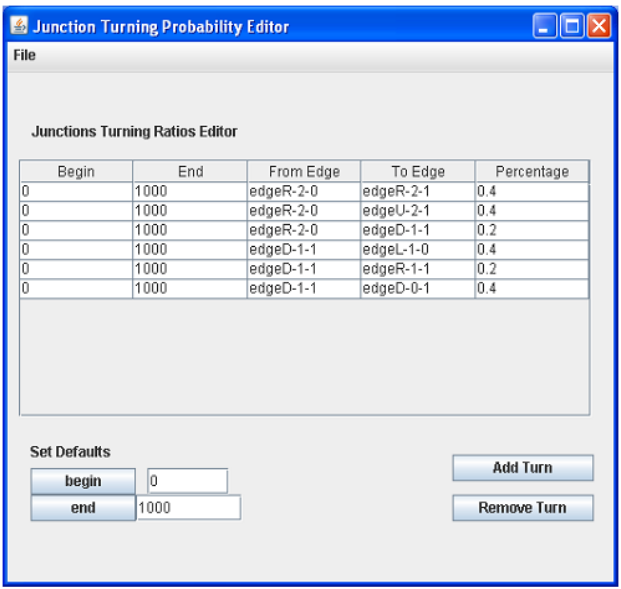

Fig. 5b

Users inputs the information in the Map Editor and the Vehicle Movement Editor then this information is fed into SUMO to generate mobility trace which can be immediately used by NS-2 version 2.34 to simulate real world vehicle movements. We can also visualize the generated mobility trace by clicking on the "Visualization" button on Main Menu, as shown in Fig, 7.

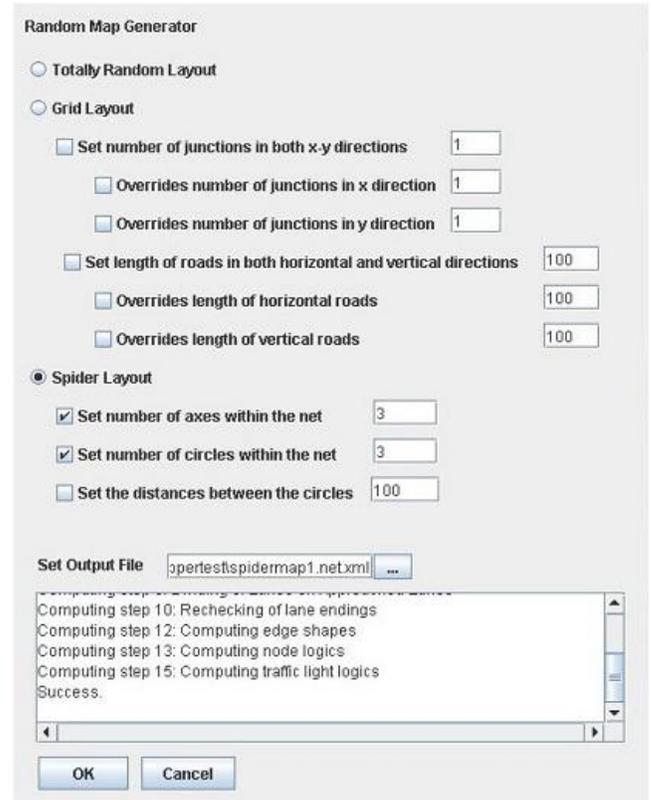

Fig. 6a

You can create a custom auto-generated map with three types of layout: grid, spider, or total random.

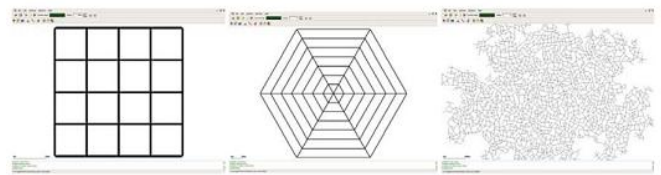

Fig. 6b

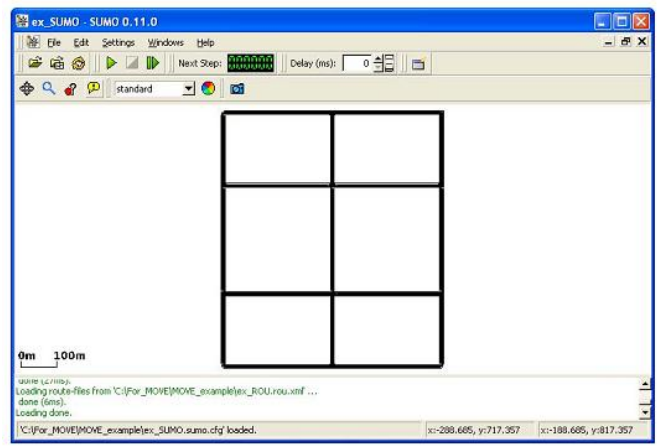

Fig. 7.

## 2.1 MAP Editor

In MOVE, the road map can be generated manually, automatically or imported from a real world map. Manual generated of the map requires inputs of two types of information, nodes and edges as shown in Fig.4, 8.

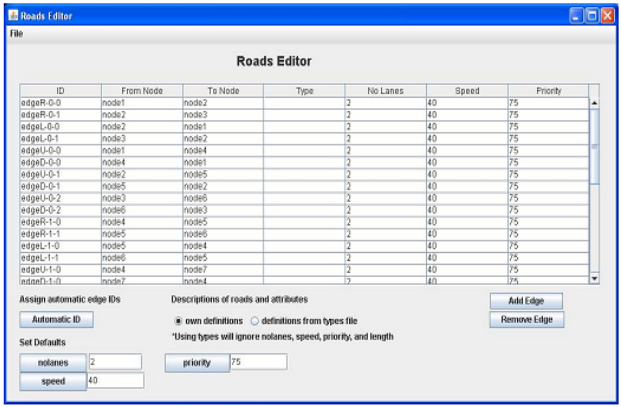

Fig. 8

A "node' Fig. 4 is one particular point on the map which can be either a junction or the dead end of the roads. Furthermore, the junction nodes can be either normal road junctions or traffic lights. The edge Fig. 8 is the road that connects two points (nodes) on a map. The attributes associated with an edge include speed limit, number of lanes, the road priority and the road length. Fig.4 and Fig.8 shows snapshots of nodes editor and edge editor.

We can integrate Google Earth into MOVE. Google Earth is a tool that enables its user to view the satellite image map of any place on earth. The functionality that Google Earth provides is called "Place mark" which allows the user to put a mark on any location of the Google earth map. Each place mark contains the longitude and latitude information of the selected locations and can be saved into a file in KML format [12]. One can define the node location on the Google map and then extract the node information by processing the saved KML file. This allows MOVE users to generate a map for any real-world road on earth for their simulations.

The road map can also be generated automatically without any user input. Three types of random maps are currently available (Fig. 6b):- grid, spider and random networks. There are some parameters associated with different types of random maps such as number of grids and the number of spider arms and circles. One can also generate a real world road map by importing real world maps from publicly available database.

## 2.2 Vehicle Movement Editor

The movements of vehicles can be generated automatically or manually using the Vehicle Movement Editor. The Vehicle Movement Editor allows users to specify several properties of vehicle routes including the number of vehicles in a particular route, vehicle departure time, origin and destination of the vehicle, duration of the trip, vehicle speed etc. We can define the probability of turning to different directions at each junction in the editor Fig. 5, 9. MOVE allows users to enter the bus time table to simulate the movements of public transport. We model the bus as one type of vehicle which has similar parameters, such as speeds, routes, etc, associated with it as other vehicles. In addition, one needs to define the departure times of the first and the last bus and the bus inter arrival time (which is assumed as constant) to simulate the bus time table Fig. 10.

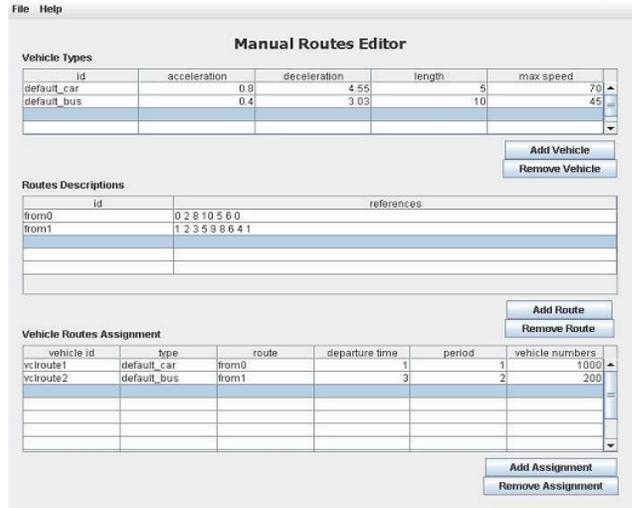

Fig. 9

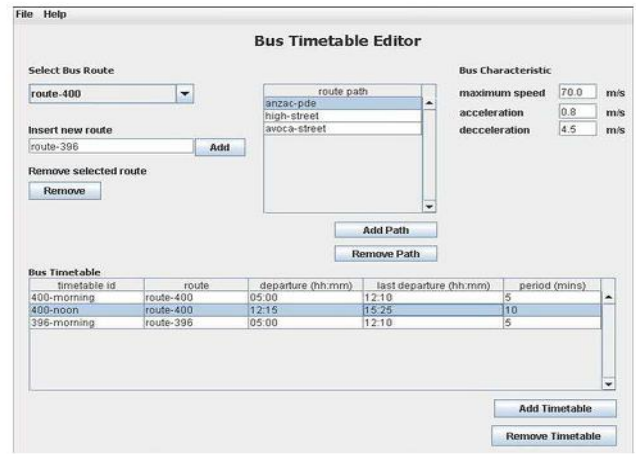

Fig. 10

## 3. EVULATION

We evaluate the impact of mobility models generated by MOVE on the performance of ad-hoc routing protocol. We compare the performance of AODV when used with the random waypoint model to that using the MOVE mobility model.

The simulation experiments were carried out in NS-2 version 2.34 on fedora 13 operating system. Each simulation lasts for 900 seconds. We generated scenarios for 150 nodes moving in an area of 4 square kilometers. The number of source nodes from 10 to 50, each of which is a CBR traffic source transmitting UDP packets of a size 64 bytes at the rate of 4 packets per second. All nodes use 802.11 MAC operating at 2Mbps. The propagation model employed in the simulation is the log normal shadowing model. We used a path loss exponent 2.56 with standard deviation 4.0 based on real world measurement data from an inter-vehicle experiment.

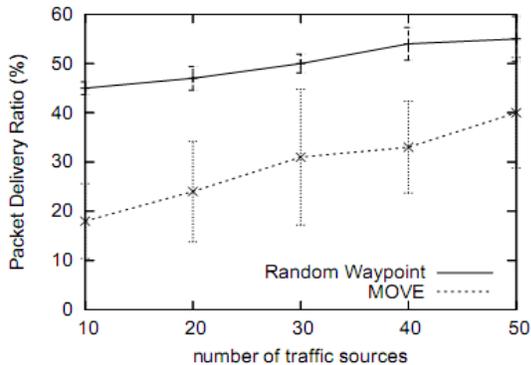

Fig. 12

Fig. 12 shows the packet delivery ratio of AODV with different number of traffic sources. Each data point represents the average of six runs and the error bars represent the range of observed packet delivery ratios. Overall, the packet delivery ratios increase as the number of traffic sources increases, which suggest a higher density of nodes can increase the network performance as long as the increasing density does not create more radio interference. The packet delivery ratios of AODV when using MOVE mobility models are lower than when using Random Waypoint model and have larger variations. The larger variance in MOVE data points is possibly due to unstable network connectivity imposed by constrained node movements by roads and traffic control mechanisms [14,15,16].

## 4. CONCULSION

In this paper, we describe the implementation and execution of a MOVE tool on Fedora 13 operating system which is based on an open source micro-traffic simulator SUMO. MOVE allows user to quickly and easily generate real world mobility models for vehicular network simulations. MOVE is publicly available and can be downloaded via following URL http://www.cs.unsw.edu.au/klan/move/. SUMO is also publicly downloaded via following URL http://sumo.sourceforge.net/. Next step is to use this tool to understand the effect of level of details in context of VANET simulation.

The movements of vehicles are based on static configurations defined in the Vehicle Movement Editor. The mobility model is first generated off-line and then used by network simulator NS-2 version 2.34.